\documentclass[11pt]{article}
\usepackage{graphicx}
\usepackage{amsmath,amssymb}
\usepackage{geometry}
\usepackage[displaymath]{lineno}
\usepackage{natbib}
\usepackage{setspace}
\usepackage{hyperref}

\title{Rupture and afterslip controlled by spontaneous local fluid flow in crustal rock}
\author{Frans M. Aben$^{1,2}$ \and Nicolas Brantut$^1$\\
  $^{1}$ Department of Earth Sciences, University College London, London, UK\\
  $^{2}$ TNO Applied Geosciences, Utrecht, The Netherlands}
\date{\ }

\begin{document}

\maketitle

\begin{abstract}
Shear rupture and fault slip in crystalline rocks like granite produce large dilation, impacting the spatiotemporal evolution of fluid pressure in the crust during the seismic cycle. To explore how fluid pressure variations are coupled to rock deformation and fault slip, we conducted laboratory experiments under upper crustal conditions while monitoring acoustic emissions and in situ fluid pressure. Our results show two separate faulting stages: initial rupture propagation, associated with large dilatancy and stabilised by local fluid pressure drops, followed by sliding on the newly formed fault, promoted by local fluid pressure recharge from the fault walls. This latter stage had not been previously recognised and can be understood as fluid-induced afterslip, co-located with the main rupture patch. Upscaling our laboratory results to the natural scale, we expect that spontaneous fault zone recharge could be responsible for early afterslip in locally dilating regions of major crustal faults, independently from large-scale fluid flow patterns.
\end{abstract}

\subsection*{Plain language summary}
Faults in rock form during what is known as rupture; rupture is followed by sliding along the newly formed fault. If this occurs quickly, we speak of an earthquake. When creating the fault during rupture, small cracks are formed in the intact rock that grow and link up. This creates local void spaces in the nascent fault zone. Pore fluids, which typically reside at pressure everywhere within the Earth's crust, will expand in this newfound space so that the pore pressure drops in the rupture zone. Such a drop may slow down the rupture and fault slip -- an earthquake may be postponed. Here, we measure this so-called dilatancy effect in unprecedented detail: We observe how the rupture is controlled by the dilatancy effect, and we discover that after rupture fluids from further away flow into the void spaces in the newly formed fault, thereby increasing pressure again and driving fault slip. This pore fluid control on fault behaviour is a fundamental mechanism that can explain why faults continue slipping just after an earthquake.

\section{Introduction}

The strength of rocks and faults in the brittle regime is in good approximation proportional to an ``effective'' stress equal to the difference between fault normal stress and the pressure of the saturating pore fluid \citep[][chap. 7]{paterson05}. This simple relationship, widely supported by observations, has many important consequences for the dynamics of faults in the Earth's crust: a rise in fluid pressure promotes faulting, while fluid pressure drops tend to inhibit it. This hydro-mechanical effect is well documented, and pore pressure increase has been demonstrated to produce seismicity in field experiments \citep[see for instance][]{raleigh76}. At crustal scale, natural seismic swarms and some aftershock sequences are interpreted as resulting from fluid flow along faults or fault systems because their spatio-temporal characteristics seem to match with those of a diffusion process, with fluid sources being typically assumed to be deep reservoirs \citep[e.g.][]{miller04,debarros20,ross20,ross21}. Such seismic sequences occur over timescales of days to years, within active regions of several kilometres, implying fluid flow on similar spatio-temporal scales. Fluctuations in pore fluid pressure are also responsible for time-dependent post-seismic deformation of the crust at km-scale, over timescales of the order of several months, via poro-elastic effects \citep[e.g.][]{peltzer98,jonsson03}.

Even in the absence of crustal-scale flow from remote fluid sources, fluid pressure variations are expected near active faults during the seismic cycle, because crustal deformation and fluid pressure variations are coupled: in low porosity, low permeability crustal rocks, failure and fault slip are associated with dilation both in the bulk, due to microcrack growth \citep[e.g.,][and many others since]{brace66}, and on the fault plane, due to overriding asperities at small (10s of $\mu$m) to large scale (10s of m) \citep[e.g.,][]{barton76,teufel81,marone90,samuelson09}. When the rate of dilation is higher than the rate of fluid flow into the fault, fluid pressure decreases \citep[sometimes violently, see][]{brantut20} and frictional strength increases \citep{brace68}. This dilatancy-hardening effect has been shown to stabilise fault slip \citep{martin80,rudnicki88,segall95,french17,aben21} and to promote slow rupture propagation \citep{rice73,segall10,brantut21b}. All these effects are transient, and pressure fluctuations may reequilibrate rapidly when deformation stops.

While the dilatancy-hardening phenomenon has been well studied in the context of rupture stabilisation, the role of fluids in the post-rupture period has not been thoroughly explored.  Recent improvements in high frequency geodetic measurements have revealed diffusion-like fault afterslip and aftershock sequences at short timescales, of the order of minutes to hours \citep[e.g.][]{jiang21}. Within such short time periods, fluid flow is expected to be spatially limited (e.g., assuming hydraulic diffusivity of $0.01$~m$^2$/s, a conservative upper bound for damaged crustal rock, the diffusion length scale is of the order of a few meters for durations of the order of 1~h). Therefore, the role of fluids during early afterslip, if any, is most likely local and should not involve crustal-scale flow. In the past few decades, afterslip has usually been explained by the rate dependency of friction on the fault plane, i.e., a viscous effect that does not rely on the existence or variations in fluid pressure \citep[e.g.][]{marone91,dieterich94,perfettini04}. Friction-based or ``viscous'' afterslip models \citep[e.g.][]{perfettini04} consider that a main shock induces an increase in shear stress on neighbouring faults, which may transiently creep provided that they obey a rate-strengthening constitutive law. This type of model predicts afterslip to be located only in the region surrounding the main shock. However, afterslip co-located with coseismic slip is a feature consistent with several observations over timescales of days to months \citep[e.g.][]{miyazaki04,zhao17}. Whether co-located afterslip occurs systematically is not yet clear, especially in the early post-seismic stage where afterslip is difficult to resolve \citep[e.g.][]{burgmann02,tsang19}.

Recent experiments reported in \citet{aben21} have revealed that fault slip in fluid-saturated rocks occurs in two stages: one accelerating phase linked to severe dilatancy and stress drop, and a prolonged decelerating phase linked to pore pressure recharge and gradual stress drop \citep[Figure 6 of][]{aben21}. This second phase can be understood as fluid-induced fault afterslip, where fluid flow originates from the fault walls into the dilated fault zone. Here, we aim to clarify the underlying physical mechanisms responsible for the macroscopically observed two-stage faulting process, and to determine the consequences for fault dynamics and possible fluid-induced afterslip in nature. In particular, the role of fault formation (i.e., propagation of a rupture in an initially intact material) vs. slip on the newly formed fault has not yet been elucidated. We combine acoustic emission locations together with mechanical data acquired at high frequency during rupture tests, and determine the exact timing of rupture propagation, fault completion (i.e., point when the fault goes through the entire sample), and slip on the newly-formed fault with respect to stress drop, pore pressure changes, and macroscopic sample deformation.

Our new observations indicate that most of the dilatant pore pressure drop occurs at the very early stages of rupture and concomitant slip, during the formation of the fault, while post-failure slip, hereafter afterslip, is mainly driven by the pore pressure recharge. We establish a simple model characterising the afterslip expected from pore pressure recharge as a function of hydro-mechanical parameters, and discuss the implications for fault dynamics in nature.

\section{Experimental approach}

To elucidate the details of fault formation versus slip in the presence of fluids, we reproduced two failure tests performed previously by \citet{aben21}: One failure test stabilised by dilatancy where the second phase of prolonged decelerating afterslip was observed, and one where dynamic rupture occurred without such a second phase. In tests by \citet{aben21} mechanical data and local on-fault pore pressure were measured during deformation. Here, we augment these measurements with active and passive ultrasonic measurements, from which we determine acoustic emission (AE) locations that delineate the rupture front during fault formation and highlight slipping parts of the newly formed fault. In addition, the mechanical and local pore pressure data are recorded at high frequency (10~kHz) during time windows in which rupture and slip occurred.

\begin{figure}
  \centering
  \includegraphics{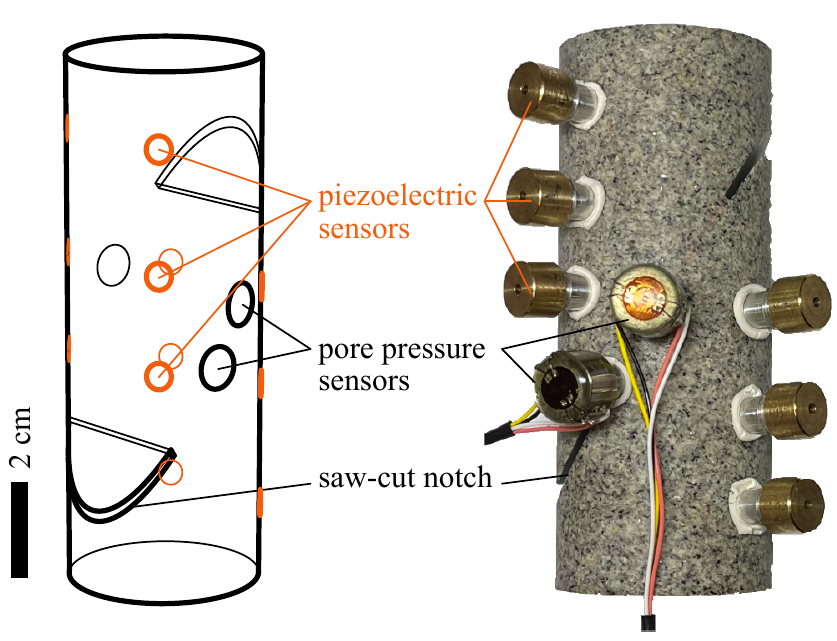}
  \caption{Schematic (left) and photograph (right) of sample geometry and instrumentation. A set of 12 piezoeletric transducers and 2 pore pressure transducers was positioned at the sample surface. The sample was deformed in the axial direction at constant confining pressure. The photograph illustrates the sensor positions and notch geometry, but for clarity does not show the Teflon spacers and nitrile jacket.}
  \label{fig:sample_sketch}
\end{figure}

We used cylindrical samples of Westerly granite, 40~mm in diameter and 100~mm in length.  To ensure that the fault forms along a predictable trajectory, the samples contained two 17 mm deep notches cut at opposite sides into the cylindrical surface at a 30$^{\circ}$ angle with the cylinder axis (Figure \ref{fig:sample_sketch}). The samples were thermally treated to induce thermal cracks and increase the hydraulic diffusivity of the material \citep[e.g.][]{darot00}. The treatment consisted in placing the samples in a tube furnace that was heated at a rate of 3$^{\circ}$C min$^{-1}$ to 600$^{\circ}$C. This temperature was maintained for the duration of two hours, followed by cooling over the course of about 12 hours by switching off the furnace. After, Teflon disks were inserted into the notches to prevent their collapse during the experiment. 

The samples were jacketed and equipped with 2 miniature pore pressure transducers \citep{brantut20,brantut21}, positioned at the rock surface directly on the prospective fault plane, and with 12 piezoelectric transducers polarised normal to the rock surface.

The two miniature pore pressure transducers that monitor on-fault pore pressure fluctuations are described in \citet{brantut21}, and consist of two parts: (1) a metal stem with a small conduit connecting a curved surface in contact with the rock to the open face at the top. The stem has a radius of 3.5~mm. An o-ring is housed in the top part of the stem; (2) a cap with a thin shoulder on the internal rim forming a penny-shaped cavity with the upper surface of the stem. The stem is sealed off by the surrounding jacket in which it is anchored with epoxy, so that fluids present only at the stem-rock interface are directly connected hydraulically to the cavity in the transducer cap. The elastic distortion of the cap caused by the pressure differential between confining pressure on the outside of the cap and the fluid pressure in the cavity is measured by a diaphragm strain gauge. The output voltage of the strain gauge is used to compute pore pressure, conditional on the separately measured confining pressure. To do so, for each new experiment the transducers are routinely calibrated during the step-wise pressurisation stage \citep{brantut21}. 

The instrumented samples were placed in the trial Rock Physics Ensemble installed in the Rock and Ice Physics Laboratory at UCL \citep{eccles05}. Mechanical data measured during the tests were confining pressure, axial load, axial shortening, local pore fluid pressure, and pore fluid pressure and pore volume in the servo-controlled pore pressure inducer. These data were recorded at 5~Hz throughout the test, and at higher frequency (10~kHz) during the window in which rupture and all fault slip occurred. Axial shortening was measured externally by a Linear Variable Differential Transformer. The sample shortening was obtained after correcting for the elastic distortion of the loading column. Corrected axial shortening was converted to ``equivalent'' fault slip, assuming the 30$^{\circ}$ fault angle and neglecting fault normal motion compared to shearing. This ``equivalent'' fault slip does not strictly correspond to actual slip when the fault is not completely through-going, and is therefore only a proxy for partial slip on the ruptured part of the sample. Axial load was measured by an external load cell, was corrected from piston seal friction and converted to differential stress. Shear stress on the plane of the prospective fault was computed from differential stress and confining pressure. Normal stress was computed from differential stress, confining pressure, and local pore pressure, and was corrected for the reduction in contact area with ``equivalent'' fault slip. Shear stress was corrected for the presence of the Teflon spacers that have a lower shear resistance than the intact rock (see Appendix A in \citet{aben21} for the full description of shear and normal stress corrections). 

Piezoelectric transducer signals were amplified to 40~dB and continuously recorded at 10~MHz and 12 bits resolution (full range $\pm5$~V). At regular intervals, the transducers were used as active sources to measure average P wave velocities in the rock \citep[see technique described in][]{brantut15}. Acoustic emission (AE) events were detected in the continuous waveforms by first extracting 40~$\mu$s time windows where at least 5 channels exceeded a threshold of 100~mV, and then using an STA/LTA criterion with a threshold of 20~dB over a 2~$\mu$s front window and 6~$\mu$s back window. The events were then autopicked using the AIC method of \citet{maeda85}, and located using a collapsing grid-search method minimising the least absolute value of arrival time errors. The P wave velocity model used for AE location was a homogeneous, elliptic anisotropic model interpolated in time from averaged P wave velocity measurements at different propagation angles with respect to the loading axis. Location errors were estimated by the product of P wave velocity and residual error in arrival times. We rejected from our analysis the events for which such errors were greater than 10~mm, or that were located outside the sample volume. In order to investigate rupture propagation, we further selected events located within a 20~mm wide region around the eventual fault plane, and projected them onto that plane (see Figure S1 and S2). The fault was then divided into 2~mm $\times$ 2~mm cells, and time series of the cumulated number of AEs in each cell were computed. The total ruptured area was estimated as the area in which AE density was greater than a threshold of 0.5~mm$^{-2}$.

The samples were saturated with water and step-wise pressurised towards the desired nominal Terzaghi effective pressure of 40~MPa, achieved by applying either a confining pressure of $P_\mathrm{c}=110$~MPa and an initial pore pressure $p_0=70$~MPa (sample WG12, which underwent stabilised failure), or a $P_\mathrm{c}=60$~MPa and $p_0=20$~MPa (sample WG15, which experienced dynamic failure). During pressurisation the permeability and storage capacity of the samples were measured as a function of pressure (see Appendix A in \citet{aben21}): At 40~MPa effective pressure the permeability was $5\times 10^{-19}$~m$^2$, and the storage capacity was $2 \times 10^{-11}$ Pa$^{-1}$. The samples were then deformed at an axial strain rate of 10$^{-6}$~s$^{-1}$, and the pore pressure was kept constant at the sample ends by a servo-controlled pore pressure intensifier. Hence, all pore pressure fluctuations measured in the sample by the local pore pressure transducers are due to partially-drained or undrained response of the material to internal pore volume changes. Deformation was stopped immediately after failure (i.e., after the main stress drop) and the sample was depressurised only after the initial pore pressure was re-established at the local pore pressure transducers.

\section{Results}

\begin{figure}
  \centering
  \includegraphics{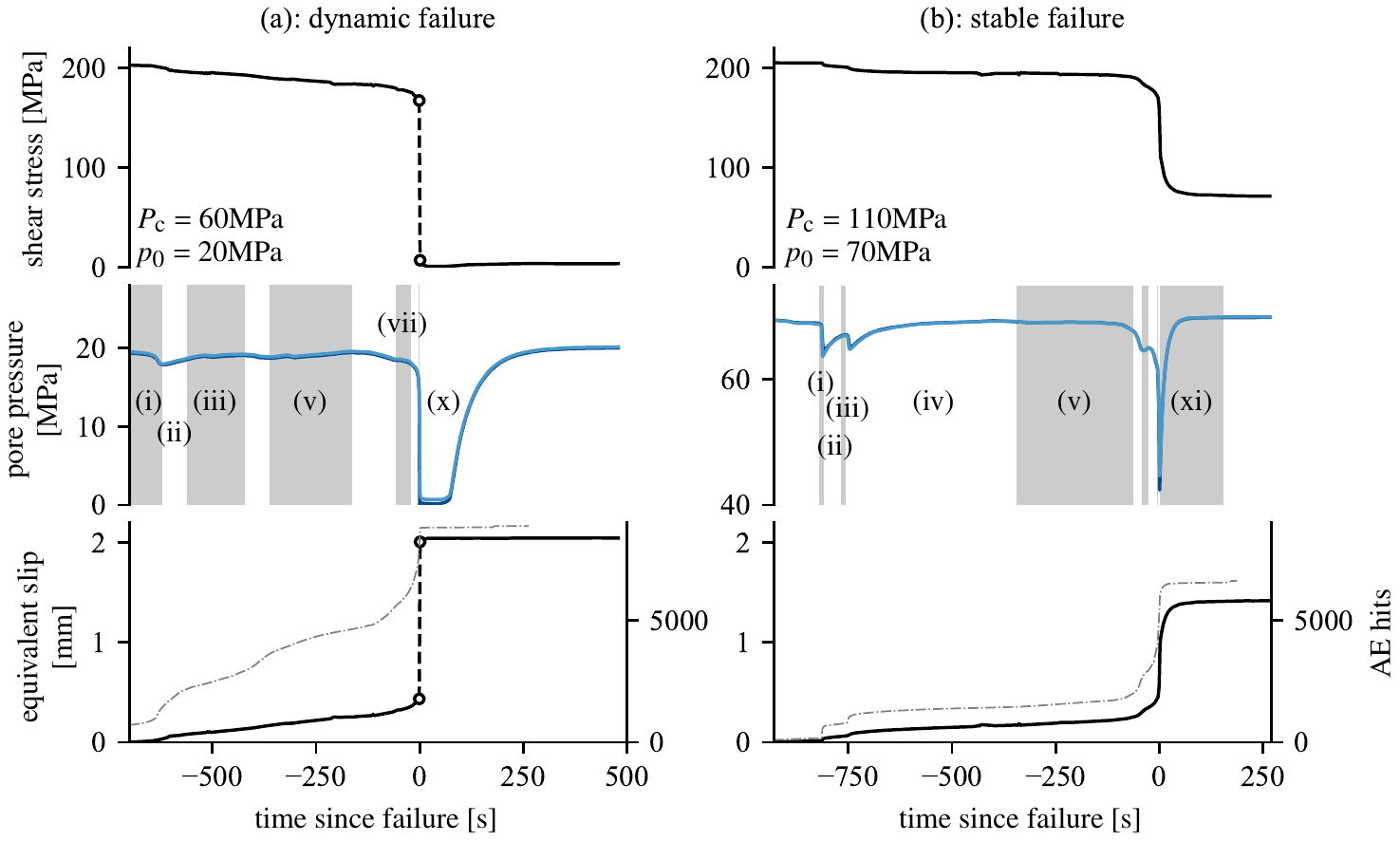}
  \caption{Time series of shear stress (top panels), local pore pressure (middle panels), equivalent fault slip (black curve, bottom panels), and cumulative AE hits (gray dashed line, bottom panels) during the entire failure and slip process for (a) sample WG15, tested at $P_\mathrm{c}=60$~MPa and $p_0=20$~MPa, and (b) sample WG12, tested at $P_\mathrm{c}=110$~MPa and $p_0=70$~MPa. Time intervals mentioned in the text are shown in gray in the middle panels, including their roman numerals. Data from the fault completion stage of sample WG12 were previously reported in \citet{aben21}.}
  \label{fig:mechtimeseries}
\end{figure}

In this section, we firstly describe the general mechanical behaviour of both samples in terms of shear stress, fault slip, pore pressure, and AE rates. To ease the description of the data, the timeseries have been subdivided into several intervals based on a change in mechanical behaviour (Figure \ref{fig:mechtimeseries}, numbered by roman numerals). From this, we identify a precursory phase which both experiments have in common, a stable main failure phase for sample WG12, tested at $P_\mathrm{c}=110$~MPa and $p_0=70$~MPa, and a dynamic failure phase for sample WG15, tested at $P_\mathrm{c}=60$~MPa and $p_0=20$~MPa. Themechanical data and concurrent AE locations of these three phases are described in more detail in the following subsections, first on the precursory stress drops, then on the stable main failure and afterslip phase, and finally on the catastrophic dynamic failure. In the last subsection we present how rupture area evolves during deformation. 

\subsection{General behaviour}

The approach to failure in both samples is characterised by small precursory stress drops of a few MPa, occurring over around 10 to 20 seconds, accompanied by concomitant on-fault pore pressure drops, slip steps, and rises in AE rate (Figure \ref{fig:mechtimeseries}a, interval (i) and \ref{fig:mechtimeseries}b, intervals (i) and (iii)). These initial events are followed by a continuous decrease in shear stress, an increase in pore pressure, equivalent slip and cumulative number of AEs (Figure \ref{fig:mechtimeseries}a: intervals (ii), (iii), (v), Figure \ref{fig:mechtimeseries}b: intervals (ii) and (iv)) -- i.e., the run-up to fault completion or a next precursory stress drop event. In the sample tested at $P_\mathrm{c}=60$~MPa and $p_0=20$~MPa, fault completion was associated with a dynamic (audible), total shear stress drop, accompanied by total fluid depressurisation and more than 1~mm slip. This occurred over a time interval of the order of a few milliseconds (see Figure \ref{fig:dynamicfailureAEs}a for a higher time resolution). Subsequently, pore pressure remained zero for around 60~s, and then gradually came back to the nominal value of 20~MPa, while no further changes in shear stress or afterslip were recorded. In the sample tested at $P_\mathrm{c}=110$~MPa and $p_0=70$~MPa, fault completion was associated with a partial shear stress drop, down to around 70~MPa, occurring over several 10s of seconds (Figure \ref{fig:mechtimeseries}b and \ref{fig:stablefailureAEs}a). We consider this a stable failure event \citep{aben21}. This stress drop occurred in two phases: a first rapid one, from 170 down to 125~MPa over about 4~s, associated with a pore pressure drop (down to 42~MPa) and equivalent slip of 0.35~mm (interval (x) in Figure \ref{fig:stablefailureAEs}a), followed by a slower drop down to 72~MPa over around 90~s (interval (xi) in Figure \ref{fig:stablefailureAEs}a), accompanied by a further 0.55~mm of afterslip and pore pressure recovery up to $p_0=70$~MPa. 

\begin{figure}
  \centering
  \includegraphics{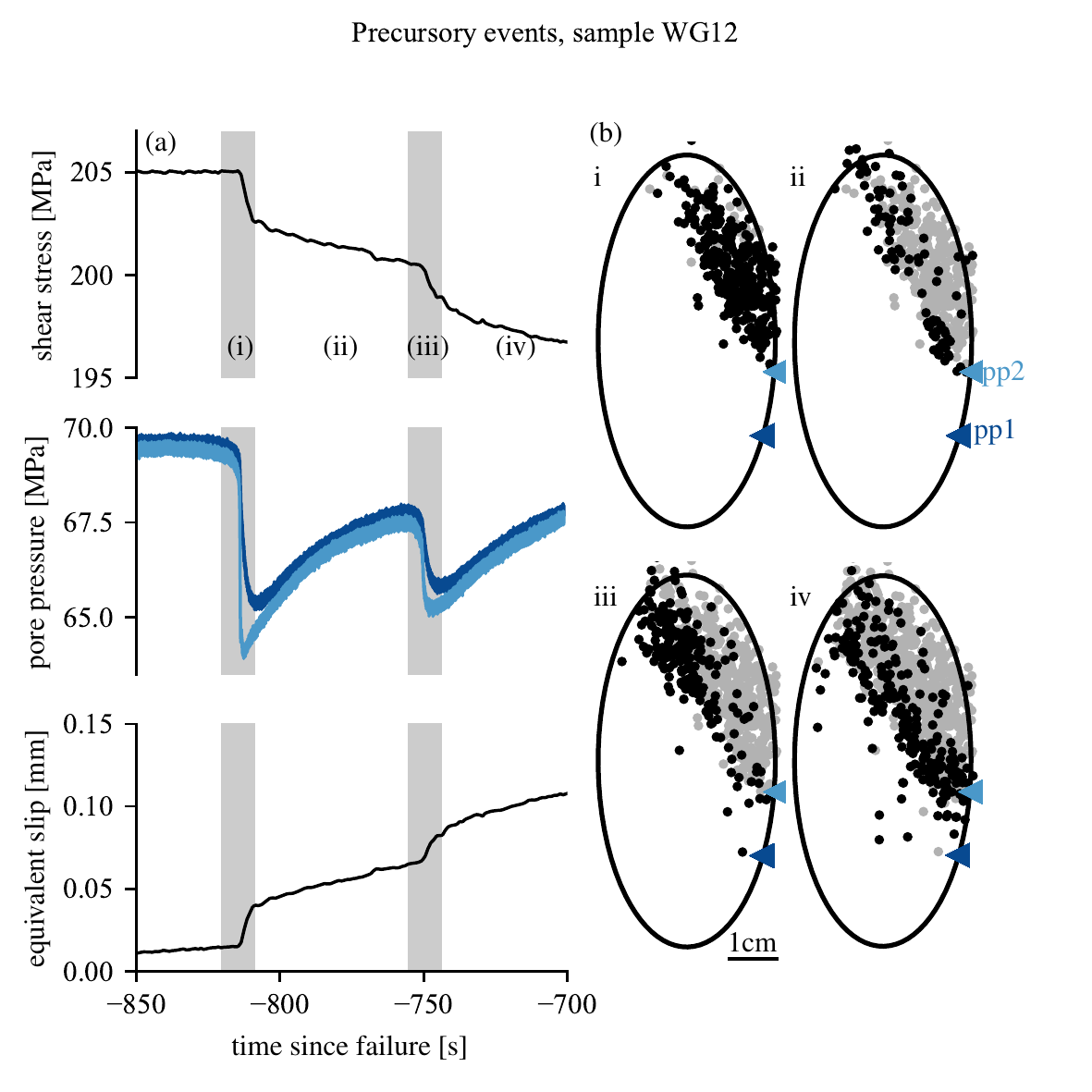}
  \caption{Mechanical (a) and AE location (b) data during precursory stress drop events (intervals labelled (i) and (iii)) in sample WG12, tested at $P_\mathrm{c}=110$~MPa and $p_0=70$~MPa. AE events are projected onto the prospective fault plane. AE events that occurred during the interval of interest are shown as black dots, and all previous events are shown as grey dots. Blue triangles labelled pp1 and pp2 correspond to locations of two pore pressure transducers along the fault plane.}
  \label{fig:precursorAEs}
\end{figure}

\subsection{Precursor stages}

In more detail, we observe that two precursory stress drops were 3 and 2~MPa, and were associated to equivalent slip increases of 25 and 15~$\mu$m, respectively (Figure \ref{fig:precursorAEs}a). They were concurrent with rapid pore pressure drops. In event (i), the local pore pressure at location pp2 (Figure \ref{fig:precursorAEs}b) dropped by 6 MPa, while at location pp1 the drop was of 4.5~MPa and appeared more gradual, with the minimum reached 6~s after the drop occurred at pp2. During phase (ii), stress continued to decrease and slip continued to increase gradually while pore pressure recovered and became uniform again. During the second event (interval labelled (iii)), pore pressure dropped again more rapidly at location pp2 than pp1. AE locations (Figure \ref{fig:precursorAEs}b) suggest that the first event (interval (i)) corresponded to partial fracture of the sample, with the fracture tip just reaching the location of pressure transducer pp2, while the other transducer pp1 was located in the unbroken region. The second event (interval (iii)) was associated with AE activity near the top of the fault plane, away from the two transducer locations. 

\begin{figure}
  \centering
  \includegraphics{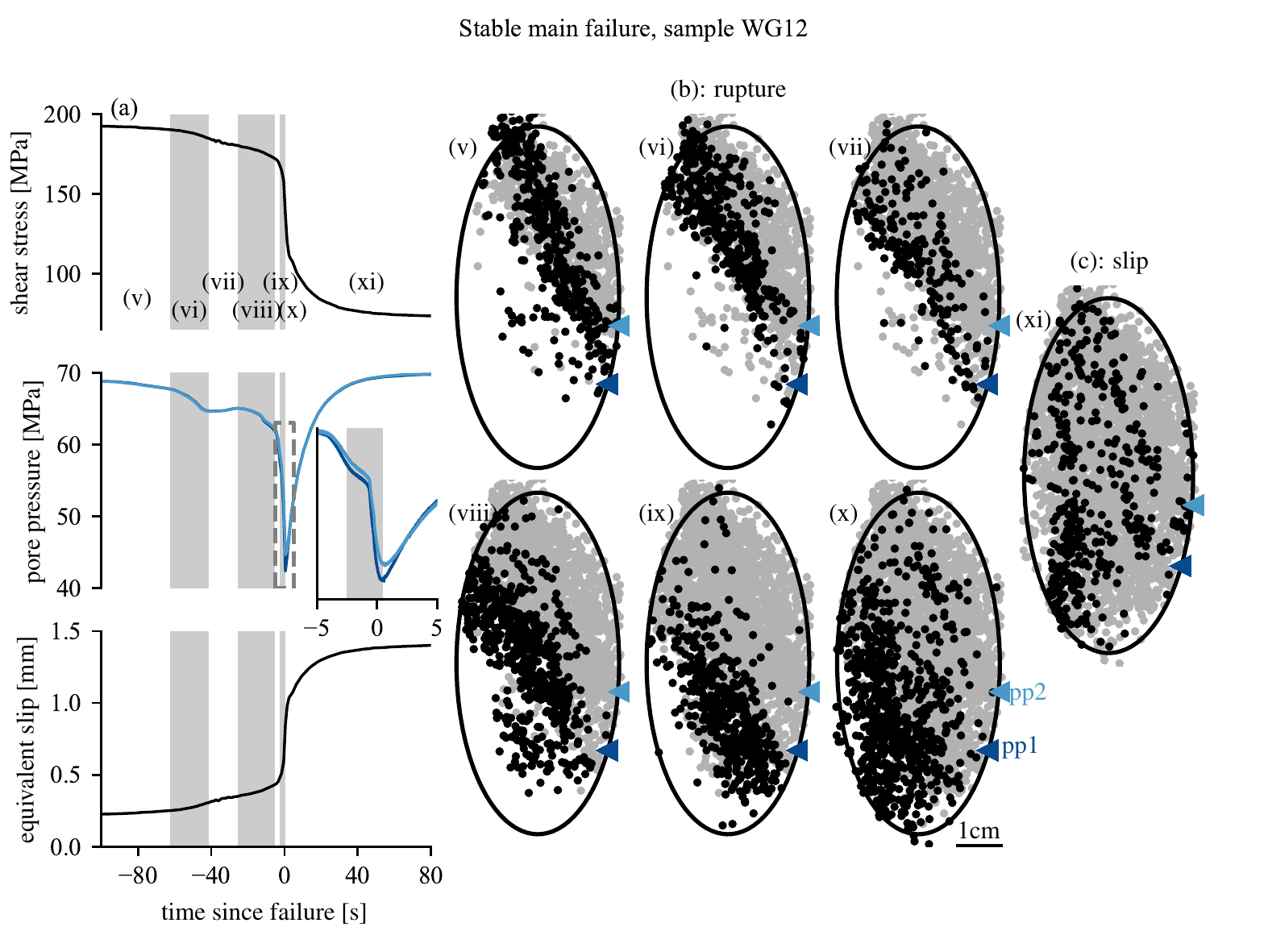}
  \caption{Mechanical (a) and AE location (b) data during main stress drop event (intervals (v) to (xi)) in sample WG12, tested at $P_\mathrm{c}=110$~MPa and $p_0=70$~MPa. See caption \ref{fig:precursorAEs}.}
  \label{fig:stablefailureAEs}
\end{figure}

\subsection{Stable failure}

The stable failure event was preceded by a gradual decrease in shear stress and increase in equivalent slip, with one phase where pore pressure dropped by around 3~MPa (Figure \ref{fig:stablefailureAEs}, interval (vi)). Throughout this preceding phase, AEs remained located in the upper part of the sample where activity was recorded previously (Figure \ref{fig:stablefailureAEs}b, intervals (v--vii)). During time interval (viii), the fault propagated further, accompanied by slip acceleration, stress decrease and pore pressure decrease. In stages (ix) and (x) (total duration of the order of a few seconds), the rupture propagated past pore pressure transducer pp1 and reached the sample's lower boundary. During these intervals, the pore pressure recorded at pp1 decreased faster and more (by a few MPa) than that at location pp2. Subsequently, AE activity became distributed again along the entire fault plane, while pore pressure recovered, shear stress relaxed and slip increased at a decreasing rate to reach a final stable value of 1.41~mm.

\begin{figure}
  \centering
  \includegraphics{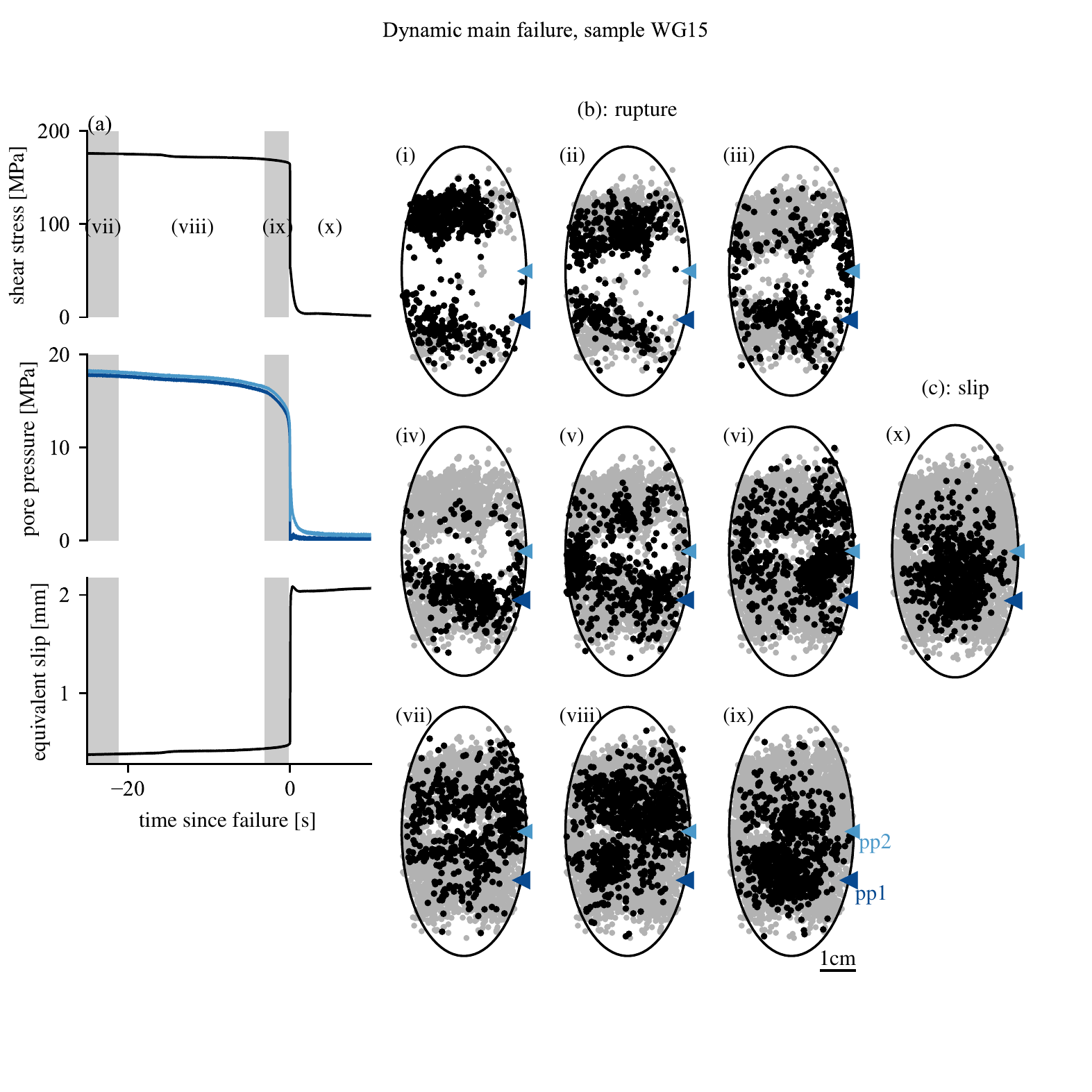}
  \caption{Mechanical (a) and AE location data during (b) and after (c) the main stress drop event in sample WG15, tested at $P_\mathrm{c}=60$~MPa and $p_0=20$~MPa. Mechanical data during intervals (i) to (vi) are shown in Figure \ref{fig:mechtimeseries}a. See caption \ref{fig:precursorAEs}.}
  \label{fig:dynamicfailureAEs}
\end{figure}

\subsection{Dynamic failure}

The sample deformed at $P_\mathrm{c}=60$~MPa and $p_0=20$~MPa experienced an early, relatively slow stress drop which coincided with the growth of partial ruptures from each notch (Figures \ref{fig:mechtimeseries}a, \ref{fig:dynamicfailureAEs}b, intervals (i) and (ii)). The rupture then gradually expanded from both ends towards the centre of the sample throughout time intervals (iii)--(viii), concomitantly with a slow, progressive stress decrease. The cumulated equivalent slip also increased by 0.32~mm. In the three seconds preceding the main stress drop (stage (ix)), rupture progressed into the central portion of the sample, while pore pressure dropped progressively and equivalent slip accelerated. Catastrophic stress and fluid pressure drop correspond to AE activity reaching the centermost part of the fault, and slip increasing by 1.57~mm. The peak slip rate was of the order of 0.25~m/s or above.

\begin{figure}
  \centering
  \includegraphics{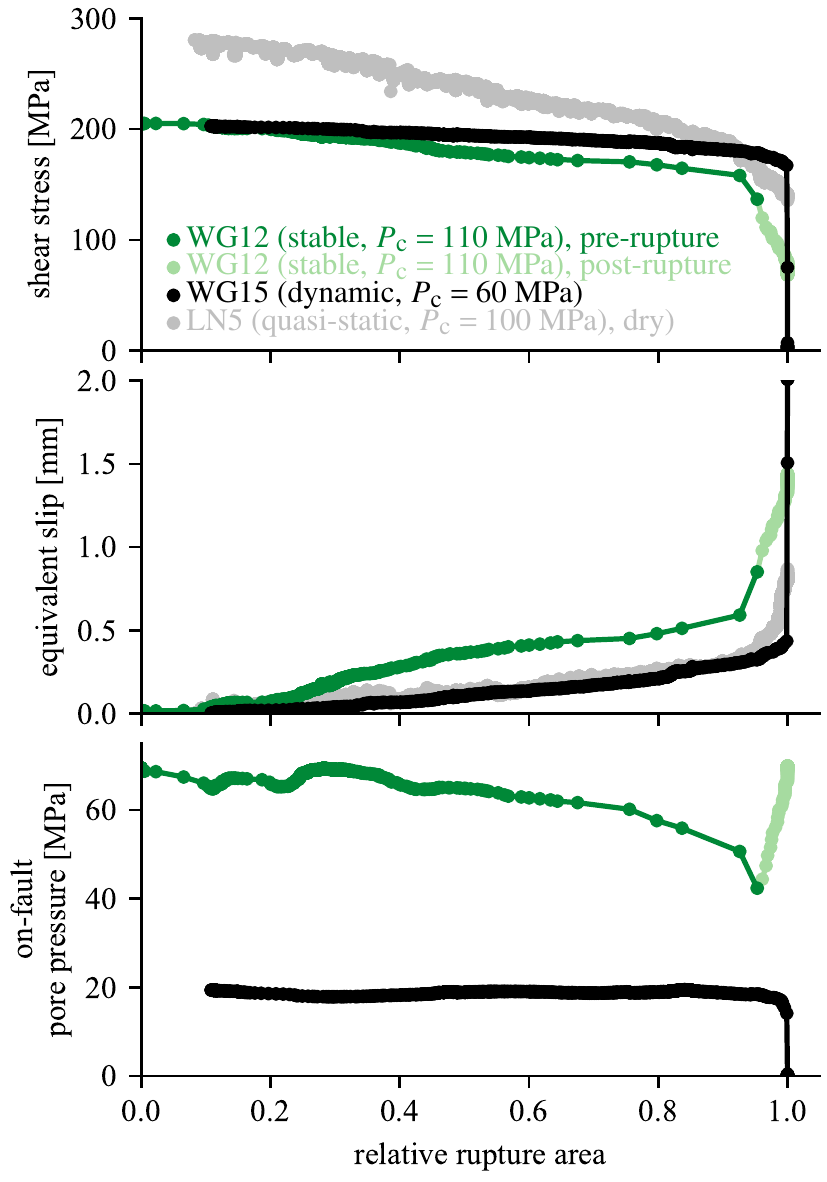}
  \caption{Shear stress, equivalent slip and local pore pressure as a function of rupture area normalised by its maximum extent, as computed from AE density projected onto the fault plane. Dots show all datapoints sampled at 1~Hz. In sample WG12, we distinguish pre- and post-rupture phases by accelerating and decelerating slip, respectively. Data from labelled LN5 correspond to quasi-static rupture growth in dry Lanh\'elin granite at $P_\mathrm{c}=100$~MPa \citep[published by ][]{aben19} and was processed in the same way as the two other datasets.}
  \label{fig:rupturearea}
\end{figure}

\subsection{Rupture area}

The calculated rupture area did not change linearly with changes in shear stress, equivalent slip and pore pressure, as computed from the AE density projected onto the fault plane (Figure \ref{fig:rupturearea}). In both tests we observed a steady, but limited decrease in stress, concomitant with a limited increase in equivalent slip, until about 90\% of the rupture was completed. For comparison, a dry quasi-static test conducted on Lanh\'elin granite at $P_\mathrm{c}=100$~MPa (Figure \ref{fig:rupturearea}, gray curve, data published by \citet{aben19}) also shows a steady, but somewhat steeper, decrease in stress up to 90\%. In the water-saturated tests, this phase was also marked by a slight, steady decrease of on-fault pore pressure, superimposed with small drops and recovery phases as the rupture area grew in a jerky manner. 

Stress drop and slip accelerated only when rupture area approached 90 to 95\% of its maximum. In the test conducted at $P_\mathrm{c}=60$~MPa and $p_0=20$~MPa (WG15, black lines in Figure \ref{fig:rupturearea}), most of the slip and stress drop was accumulated as rupture reached completion (i.e., at 99 to 100\%). This was also the case in the dry quasistatic test, although the stress drop was more limited. By contrast, in the test conducted at $P_\mathrm{c}=110$~MPa and $p_0=70$~MPa (WG12, green lines in Figure \ref{fig:rupturearea}), the stress drop and slip accelerated when a minimum pore pressure was reached, at around 95\% of the relative rupture area. From the spatial distribution of AEs we infer that at this 95\% interval the fault has been completed (Figure \ref{fig:stablefailureAEs}b). Stress drop and slip then continued into a decelerating phase while pore pressure gradually recovered. In this phase, the last 5\% of the relative rupture area metric stems from AEs highlighting parts of the fault interface where not many AEs were recorded during fault formation (Figure \ref{fig:stablefailureAEs}c). In this sample, most of the slip occurred in this latter phase, when rupture was already complete (i.e., a through-going fault was fully formed). Hence, we label such slip after fault completion as afterslip. 

\section{Discussion}

Overall, AE location analysis coupled with pore pressure and mechanical data demonstrate that rupture of water-saturated, initially intact rocks occurs in two distinct phases: (1) a rupture phase, where a fault is formed in the initially intact rock, in association with small slip and strong dilatancy (pore pressure drops); (2) an afterslip phase, where displacement occurs on the newly formed fault, in association with distributed AE activity on the fault, and a net increase in fluid pressure. What are the mechanisms driving each phase and what do they imply for the dynamics of faulting in nature?

\subsection{Phase 1: Rupture growth and dilatancy}

The formation of a shear failure plane is accompanied by an increase in sample compliance and therefore by an increase in shortening. Rupture growth is not always steady but can occur in bursts: The AE location results show that partial stress drops, accompanied by local pore pressure drops and steps in sample shortening are linked to rapid propagation and arrest of the main shear fault, which does not reach the full extent of the sample (Figure \ref{fig:precursorAEs}). During those partial ruptures, pore pressure drops more rapidly in regions overlapping with the ruptured area, which indicates that (1) dilatancy is co-located with the shear fault, and (2) only minute amounts of slip lead to significant dilatancy. As the main failure event is approached, on-fault pore pressure drops become large, and are directly correlated with increases in slip. This further reinforces the idea that dilatancy is induced by shear deformation on the fault plane.

Our observations are consistent with the micromechanics of faulting in low porosity rocks \citep{wong82b,lockner91,zang00}: The early stages of rupture propagation correspond to the formation of a connected network of cracks. This network of cracks is likely tortuous and the newly formed fault is expected to be rough: initial slip on the fault therefore leads to strong dilation.

Our results also show that only a fraction of the stress drop, slip and pore pressure changes occur before the completion of rupture propagation across the sample. In all our experiments, the largest proportion of the total fault slip was accumulated as the rupture area was near completion (Figure \ref{fig:rupturearea}). There is some variability in the stress drop and slip evolution with increasing fault area (Figure \ref{fig:rupturearea}). This could also be observed in previous AE location studies by \citet{lockner92}. While the effective compliance of the sample should scale with fault length and with the ratio of stress drop to shear modulus of the rock, the quantitative evolution of sample shortening with rupture growth depends on the orientation of fault segments with respect to the applied loads. For instance, subvertical rupture growth leads only to small changes in axial compliance, which is consistent with the rupture patterns observed in sample WG15 (Figures \ref{fig:dynamicfailureAEs} and \ref{fig:rupturearea}). Therefore, the variability observed in stress and slip evolution with rupture area can be attributed to variability in fracture geometry, which is eventually controlled by the details of sample heterogeneity and loading conditions. This variability is probably at the source of the difference in rupture growth between the two samples tested here, with one rupture (in WG12) originating at one end and propagating unilaterally, and the other (WG15) originating from both notches.

The shear failure plane propagates in bursts until a small ligament of intact rock is left in the sample. Then, stress drop and shortening accelerate dramatically as this ligament is broken. When fluid pressure is sufficiently high, dilatancy hardening prevents acceleration of motion to dynamic levels. Slip stabilisation is possible because dilation is particularly strong at small slip, as the rock loses cohesion \citep[][]{aben21}. Slip on the newly formed fault decelerates as the applied stress equates the frictional strength of the rock (Figure \ref{fig:mohrplot}). This is when the pore pressure on the fault reaches a minimum (Figure \ref{fig:stablefailureAEs}a, transition from interval (x) to (xi)), and it coincides with rupture completion, i.e., a through-going fault now exists in the sample.

\subsection{Phase 2: Afterslip due to pore pressure recharge}

The sample that experienced dynamic rupture showed a total pore pressure drop, a total stress drop and no afterslip. The pore pressure drop to zero implies that the fault is effectively ``dry'' during failure, which is due to a dilatancy effect that exceeds the initial pore pressure \citep{brantut20}. The observation of total stress drop is not uncommon \citep[e.g.][]{lockner17} and can be attributed to a combination of inertia of the loading piston, dramatic transient weakening of the fault, and loss of control of the testing machine. In our setup, all tests conducted on dry granite lead to total stress drops and no afterslip, consistently with the observation reported here in sample WG15 that experienced a full pore pressure drop.

\begin{figure}
  \centering
  \includegraphics{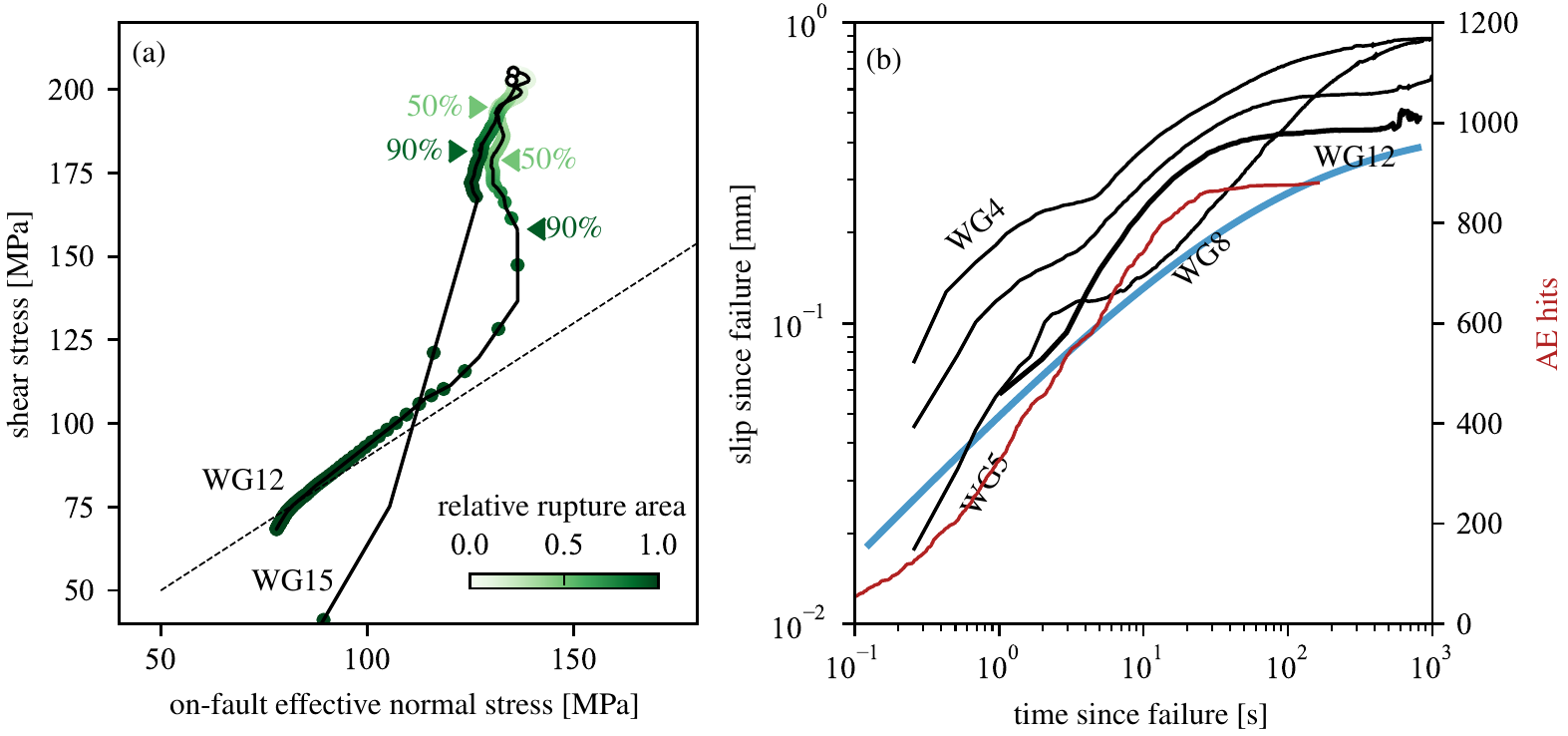}
  \caption{(a) Stress trajectories for fracture tests, colour-coded by relative fracture area as measured from AE density. (b) Afterslip as a function of time, measured from the instant when pore pressure reaches its minimum. Sample numbers refer to data reported in \citet{aben21}. The cumulative number of AE hits for sample WG12 is shown in red. The blue line is the modelled afterslip using Equation \eqref{eq:d_after}, with parameters reported in the main text.}
  \label{fig:mohrplot}
\end{figure}

By contrast, in the sample that experienced rupture stabilisation, significant afterslip could be accumulated after the main stress drop, with AE activity mostly located in the previously ruptured part, while pore pressure recovered on the fault. During this phase, afterslip is directly proportional to pore pressure recovery \citep[][and Figure 5]{aben21}, which indicates frictional slip on the newly formed fault (Figure \ref{fig:mohrplot}a). We also note that the cumulative number of AEs is directly proportional to the amount of afterslip (Figure \ref{fig:mohrplot}b), which implies that AE activity illuminates the on-fault slip process, consistently with previous observations \citep[e.g.][]{goebel12}.

Considering that (1) the fault obeys the effective stress law (as evidenced by the linear stress path once the fault is formed in Figure \ref{fig:mohrplot}a), and (2) slip on the fault is linked to shear stress drop via the machine stiffness, we establish that the pore pressure increase on the fault is the direct cause of afterslip. We can estimate the afterslip associated with pore pressure reequilibration after an initial on-fault pore pressure drop using a spring-slider model. We assume a planar fault of finite width $w$ and storativity $S_\mathrm{f}$ embedded in a host rock of diffusivity $c_\mathrm{hy}$ and storativity $S$. We impose an initial pore pressure drop of $\Delta p_\mathrm{u}$ in the fault, and compute the time-evolution of pore pressure assuming diffusion perpendicular to the fault plane. The fluid pressure $p$ is governed by a diffusion equation,
\begin{linenomath}
  \begin{equation}
    \frac{\partial p}{\partial t} = c_\mathrm{hy}\frac{\partial^2 p}{\partial y^2},
  \end{equation}
\end{linenomath}
where $y$ is the coordinate normal to the fault plane, and $c_\mathrm{hy}$ is the hydraulic diffusivity of the bulk. The boundary condition at large distances from the fault (positioned at $y=0$) is that of a constant reference pore pressure $p_0$. We consider that fluid pressure is uniform within the fault due to elevated permeability, so that the fault acts as a closed reservoir located at $y=0$. The boundary condition there is derived from the balance between the rate of fluid storage within the fault, the flux at the fault boundary, and the volume change within the fault that produces the initial pore pressure change, which leads to:
\begin{linenomath}
  \begin{equation}
    \frac{\partial p}{\partial t} - \frac{2c_\mathrm{hy}}{wS_\mathrm{f}/S}\frac{\partial p}{\partial y} = -\Delta p_\mathrm{u}\delta_\mathrm{Dirac}(t).
  \end{equation}
\end{linenomath}
The resulting pore pressure evolution within the fault ($y=0$) is given by \citep[][Chap. 13]{carslaw59}
\begin{linenomath}
  \begin{equation}
    p(t) = (p_0 - \Delta p_\mathrm{u})e^{t/t\mathrm{diff}}\mathrm{erfc}(\sqrt{t/t_\mathrm{diff}}),
  \end{equation}
\end{linenomath}
where $t_\mathrm{diff}$ is the characteristic fault recharge time, expressed as $t_\mathrm{diff}=(wS_\mathrm{f}/S)^2/4c_\mathrm{hy}$. Note that our analysis neglects the finite dimension $L$ of the sample, which is valid for timescales that are small compared to $L^2/c_\mathrm{hy}$.

Static equilibrium between the stress imposed by the elastic loading system and the frictional strength of the fault leads to the following constraint on fault slip $\delta$:
\begin{linenomath}
  \begin{equation}
    k(\delta_\infty - \delta) = \mu(\sigma - p),
  \end{equation}
\end{linenomath}
where $k$ is the spring stiffness, $\delta_\infty$ is the remotely applied displacement, $\mu$ is the friction coefficient, and $\sigma$ is the normal stress on the fault. Assuming that the initial slip $\delta(0)$ after phase 1 (fracture completion) is at equilibrium with the undrained pore pressure $p(0) = p_0 - \Delta p_\mathrm{u}$, the afterslip is given by \citep{aben21}
\begin{linenomath}
  \begin{equation} \label{eq:d_after}
    \delta_\mathrm{after}(t) = \delta(t) - \delta(0) = \frac{\mu}{k}\Delta p_\mathrm{u}\big(1 -e^{t/t\mathrm{diff}}\mathrm{erfc}(\sqrt{t/t_\mathrm{diff}})\big).
  \end{equation}
\end{linenomath}

The diffusion time is difficult to estimate: our small-scale experiments indicate a lower bound of the order of tens of seconds to a few minutes, but it could be orders of magnitude larger if a wide region experiences dilation (large value of $w$; see Figure \ref{fig:magnitudes}a). When $t\gg t_\mathrm{diff}$, Equation \eqref{eq:d_after} is usefully simplified as
\begin{linenomath}
  \begin{equation} \label{eq:d_after_simple}
    \delta_\mathrm{after}(t) \approx \frac{\mu}{k}\Delta p_\mathrm{u}\left(1 -\frac{1}{\sqrt{\pi t/t_\mathrm{diff}}}\right)\qquad (t\gg t_\mathrm{diff}).
  \end{equation}
\end{linenomath}

The modelled afterslip vs. time evolution agrees qualitatively well with our dataset and that reported in \citet{aben21} (Figure \ref{fig:mohrplot}b), using independently measured effective triaxial stiffness $k=243$~GPa/m, friction $\mu=0.63$, undrained pore pressure $\Delta p_\mathrm{u}=30$~MPa, bulk storativity $S=10^{-11}$~/Pa, fault storativity $wS_\mathrm{f}=10^{-12}$~m/Pa, and bulk diffusivity $c_\mathrm{hy}=2.5\times10^{-5}$~m$^2$/s \citep{brantut21}. The model also explains why the sample that experienced dynamic rupture with total stress drop did not show any afterslip, despite the time-dependent pore pressure recovery: since the stress drop is total due to dynamic overshoot (probably linked to piston inertia), there is no further driving force to produce afterslip.

Quantitative differences between measured and model predictions and variability between samples can be attributed to the geometrical details of each rupture, which impact the flow paths for the fluid to reach the fault plane, and to the strong simplification of the model which assumes 1D flow perpendicular to the fault plane, and neglects the presence of constant pore pressure boundary conditions at both ends of the samples \citep{aben21}. The model also neglects any further dilatancy or compaction that could occur on the fault during the afterslip phase: those effects cannot be resolved by our experimental data that only record the net pore pressure change.

\subsection{Implications and upscaling}

In our laboratory samples, we observe sequentially the rupture (fault formation) phase and a subsequent afterslip phase. This separation in time arises from the geometry of the test and the finiteness of the rock sample: due to the stiffness of the unbroken rock, only limited amounts of slip occur if the fault is not through-going.

During large scale faulting in nature, this separation in time is expected to translate also into separation in space, with the rupture tip being the region concentrating dilation and pore pressure drops, while regions far behind are subjected to pore pressure recharge and prolonged afterslip \citep{rice73,brantut21b}, even after rupture has ceased to propagate.

Our observations conducted in initially intact rock are directly relevant to rupture and slip in ``immature'' fault systems, or faults that have been substantially sealed or healed during the interseismic period. In more ``mature'' fault rocks containing clay-rich gouges, dilatancy is also significant \citep[e.g.,][]{ashman23}, and we expect a similar qualitative behaviour to that observed in intact rocks, albeit with different magnitudes for fluid pressure drops.

\begin{figure}
  \centering
  \includegraphics{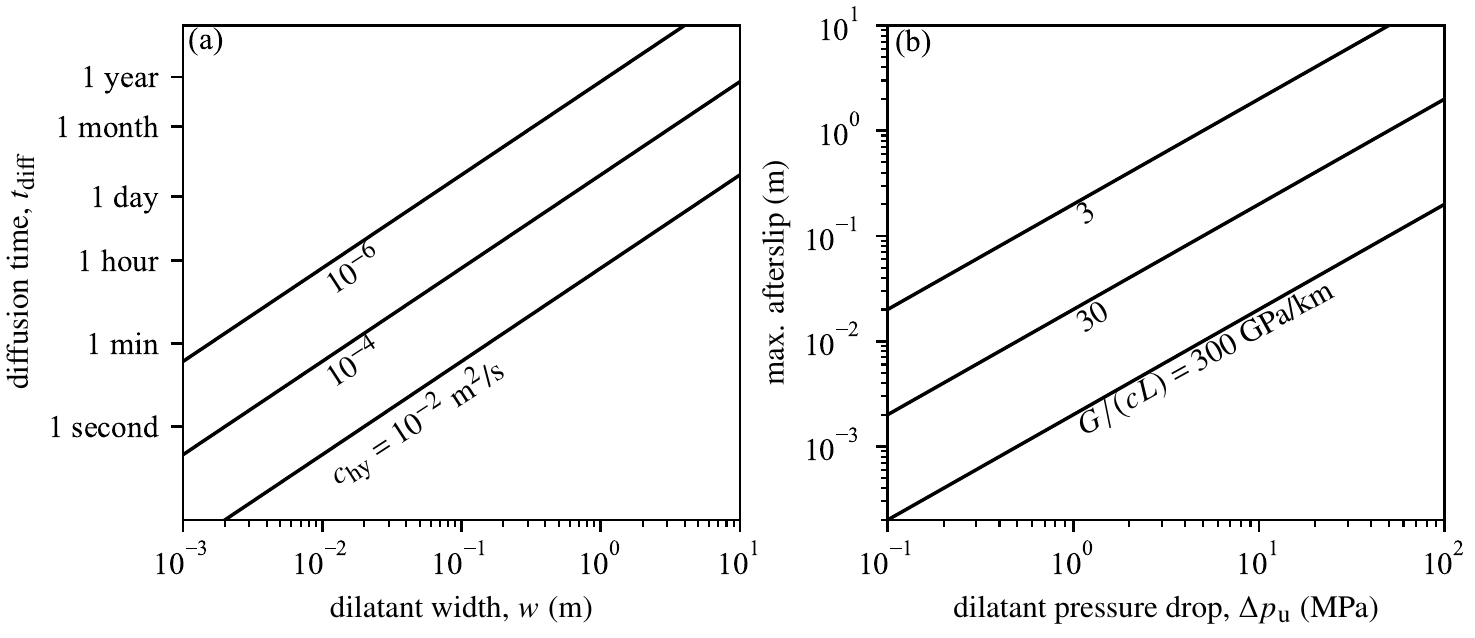}
  \caption{(a) Characteristic diffusion time computed using $S_\mathrm{f}/S = 10$ as a function of dilatant fault width, for a range of off-fault hydraulic diffusivities. (b) Maximum amplitude of afterslip due to pore pressure recharge as a function of undrained co-seismic pore pressure drop, using a friction coefficient of $\mu=0.6$, and for a range of fault equivalent stiffnesses.}
  \label{fig:magnitudes}
\end{figure}

Our observation of early afterslip due to pore pressure recharge is consistent with a simple diffusion model coupled to elastic relaxation of the surrounding material. In the laboratory, this role is played by the loading piston with stiffness $k$. In nature, the stiffness of the material surrounding a fault is related to the fault size. Assuming a uniform stress drop on a finite fault of lateral extend $L$, the equivalent stiffness relating the average slip on the fault to the stress drop is given by $G/(cL)$, where $G$ is the shear modulus of the surrounding medium and $c$ is a nondimensional factor of order 1 that accounts for the shape of the fault \citep[e.g.][]{madariaga09}. In practice, how much afterslip can be produced by fluid recharge and over what timescales depend on the hydromechanical properties of the fault and its surrounding materials (see Equation \eqref{eq:d_after}), as well as the possible stress over- or undershoot. There is considerable uncertainty in possible values for the coseismic pore pressure drop $\Delta p_\mathrm{u}$, which sets the amplitude, and the width $w$ over which the pressure drop occurs, which sets the timescale (Figure \ref{fig:magnitudes}). For relatively ``thick'' faults, say $w$ of around 10~cm, in low permeability host rocks such a crystalline basements ($c_\mathrm{hy}=10^{-5}$~m$^2$/s), we expect pore pressure recharge to occur over several hours to a day. Assuming no overshoot, the amount of afterslip is directly proportional to the pore pressure drop and to the inverse of the fault length: for $L=1$~km and $G=30$~GPa, reequilibration from a pore pressure drop of a few MPa is expected to produce afterslip of the order of few 10s of cm.

 Another quantity of interest is the ratio of afterslip $\delta_\mathrm{after}$ to total slip accrued during the entire rupture process $\delta_\mathrm{final}$. In our test, direct measurements yield $\delta_\mathrm{after}/\delta_\mathrm{final} = 0.42$. In a spring-slider model, or, equivalently, for uniform slip on a finite fault model, this can be estimated as
\begin{linenomath}
  \begin{equation}
    \delta_\mathrm{after}/\delta_\mathrm{final} = \mu \Delta p_\mathrm{u}/\Delta\tau_\mathrm{final},
  \end{equation}
\end{linenomath}
where $\Delta\tau_\mathrm{final}$ is the stress drop that is eventually observed under drained conditions, after complete reequilibration of pore pressure with the surrounding rock. Seismological data indicate earthquake stress drops of the order of 1 to 100~MPa, which implies that the afterslip associated to fluid pressure reequilibration could be a large fraction of the total slip if local dilatancy-induced pore pressure drops are of a few MPa, which is not unlikely.

Based on the above considerations, we expect near-fault fluid pressure fluctuations to play a significant role in the spatio-temporal evolution of early afterslip. Afterslip and aftershocks are commonly observed over a wide range of timescales, up to several years following earthquakes, in regions neighbouring the main rupture patch \citep[][Section 5.2.3]{scholz19}. We expect afterslip due to pore pressure recharge to be significant in the early stages following coseismic rupture, typically from minutes to hours, depending on local permeability and fault width. Such early afterslip remains technically challenging to observe, but recent seismological and geodetic analyses have demonstrated that it can be significant, starting in the first few minutes after the mainshock \citep{tsang19,jiang21,twardzik21}. One additional feature of the pore pressure recharge model is that afterslip is co-located with the main rupture patch, which is not predicted by rate-and-state afterslip models \citep{marone91,perfettini04}. Overlapping regions of co-seismic slip and early afterslip are however sometimes observed \citep[e.g.][]{miyazaki08,tsang19,jiang21,twardzik21}, which the recharge model could explain. Early afterslip on the main rupture patch is expected to amplify stress redistribution in the surrounding region, and is therefore not incompatible with other afterslip processes and typical aftershock distributions.

\section{Conclusion}

Laboratory rock deformation experiments in water-saturated granite instrumented with in-situ fluid pressure sensors and imaged by acoustic emission locations show that shear rupture occurs in distinct stages: (1) a precursory stage marked by intermittent fracture growth with limited sample deformation, (2) accelerating deformation and large pore pressure drops at rupture completion, and (3) early afterslip due to pore pressure recharge from the fault walls. All these stages are controlled by local fluid pressure changes and flow, and therefore correspond to internal dynamics of faults and their immediate surroundings, independently from crustal-scale fluid migration that might occur at longer timescales \citep[e.g.][]{miller04,ross20}. Such internal dynamics might be responsible for significant early afterslip, which is now detectable thanks to high frequency geodetic measurements \citep[e.g.][]{jiang21}.

\paragraph{Acknowledgments} This work was supported by the UK Natural Environmental Research Council grant NE/S000852/1 and the European Research Council under the European Union's Horizon 2020 research and innovation programme (project RockDEaF, grant agreement \#804685). We thank Manon Dalaison, Romain Jolivet and Phil Meredith for helpful discussions. Comments by J\"org Renner, John Rudnicki, the Associate Editor and two anonymous reviewers helped improve the manuscript.

\paragraph{Open research statement}
All data needed to evaluate the conclusions in the paper can be found at the NGDC repository of the British Geological Survey \\
(\href{https://webapps.bgs.ac.uk/services/ngdc/accessions/index.html}{https://webapps.bgs.ac.uk/services/ngdc/accessions/index.html}, where data for samples WG12 and WG15 are found under ID numbers: 165485 and 176667, and data for sample LN5 under ID number 128186) \citep{aben23}.

\bibliographystyle{agufull}
\bibliography{localbib}

\end{document}